\documentclass[12pt]{article}

\usepackage{amssymb,graphicx,epsf}
\begin{document}

\title{\textbf{Multi transitions in MgB$_{2}$ films prepared by pulsed laser deposition}}

\author{Yi Zhao\footnote{\emph{Email Address:} yzhao@physics.utoronto.ca (Yi Zhao)}, Yuemeng Chi, Ruijuan Nie, Furen Wang
\\ \emph{State Key Laboratory for Mesoscopic Physics,}
\\ \emph{School of Physics, Peking University, Beijing, 100871}}
\date{}

\maketitle

\begin{abstract}
We have grown MgB$_{2}$ films in a postannealing process.
Precursors were prepared on Al$_{2}$O$_{3}$(0001) substrates by
codeposition of Mg and MgB$_{2}$ using pulsed laser deposition
technique. Superconducting MgB$_{2}$ thin films were obtained via
an \emph{ex situ} postannealing process, with various annealing
temperatures and durations. In magnetic measurement we found more
than one transitions in almost all the samples, and the
temperature dependence of the resistance confirmed this phenomena.
We proved this is not result of magnesium deficiency.
Transportation properties of MgB$_{2}$ thin films under strong
magnetic fields are also studied.
\\\emph{Key words:} superconducting, MgB$_{2}$, film, annealing
\\\emph{PACC:} 6855, 7470L, 7475, 8115I
\end{abstract}

\section{Introduction}

 The discovery of superconductivity in MgB$_{2}$
compound \cite{ref1}, with the highest transition temperature
around 40K\cite{WangSF2001}\cite{Wang2002}, has aroused great
interest on investgating the properties of this promising
material. The strongly linked nature of the
inter-grains\cite{dcl2001} with high charge carrier density and
the relatively simple structure\cite{Zhang2001}\cite{He2001} of
this material reveals its perspective for use in technological
applications. Moreover, as a two-band superconductor, its simple
crystal structure and high T$_{c}$ make it a perfect and real
system to study the fundamental properties of two-band
superconductivity\cite{pickett2002}\cite{choi2002}.
\par
There are two approaches of the postannealing technique: the
\emph{in situ} method and the \emph{ex situ} method. The \emph{in
situ} process is more compact and more likely to grow smooth
films, however, this way of growth has several disadvantages, like
poor crystallinity, multiphase, and relatively low
T$_{c}$\cite{HMC2001}\cite{Ivanov2002}. \emph{Ex situ} methods are
exploited to overcome both of the two essential factors for the
fabrication of superconducting MgB$_{2}$ thin films: the vapor
pressure of magnesium and the oxidation of magnesium, and have
achieved films with quite high T$_{c}$ and
J$_{c}$\cite{wnk2002}\cite{wnk2003}. Recently the \emph{ex situ}
approach has been greatly improved. As an essential part of
superconductor device, Josephson junction has been fabricated on
\emph{ex situ} grown MgB$_{2}$ thin films\cite{Ivanov2004}.
\par
In this paper, we report the superconducting properties, including
some unusual phenomena, of MgB$_{2}$ thin films fabricated under
various annealing temperatures and durations. In magnetic
measurement we found more than one transitions in almost all the
samples, and the temperature dependence of resistance confirmed
this phenomena. We compared our results with some other groups'
works, and proved a different explanation. We also studied the
transportation properties of samples under different magnetic
fields, extracting their upper critical fields.

\section{Experimental}
The MgB$_{2}$ thin films used in this study were prepared in an
\emph{ex situ} postannealing approach: the precursors were all
prepared on Al$_{2}$O$_{3}$(0001) substrates by codeposition of
high purity Mg and MgB$_{2}$ targets at room temperature using
pulsed laser deposition technique in a high vacuum condition of
$\sim1.5\times10^{-7}$ Torr. The laser energy and pulse repetition
rate were 700mJ/pulse and 8 Hz, respectively. An additional Mg
layer was deposited as a cap on the precursor immediately after
codeposition. The precursors were black with metal luster, and
darkened after being exposed to the air for a few days, which is
believed to be the oxidation of magnesium coat.
\par
After deposition, the precursors were put in a sealed Ta tube and
annealed in Mg atmosphere. The annealing temperatures were
$600^{\rm o}$C, 700$^{\rm o}$C, 800$^{\rm o}$C, 900$^{\rm o}$C and
1000$^{\rm o}$C, respectively. The durations of the annealing
process were 10min, 20min, 40min and 60min, respectively. Pure
argon was flowing around the samples during the thermal treatment.
Enough magnesium tapes were applied to keep a high Mg vapor
pressure. Special measures were taken to make sure that the
magnesium tapes would not contact the surface of samples directly.
The samples after annealing were either golden or black.
\par
The magnetization versus temperature measurements were performed
using standard SQUID magnetometer device (quantum design MPMS).
Sample was first cooled to 5K in zero field, then a low magnetic
field of 50Oe was utilized in the direction parallel to the
surface of the sample. The sample was then heated to 50K in this
field. The magnetization of sample was measured every 1K after the
stablization of system temperature. Measurements of magnetic
properties were performed on all samples.
\par
We obtain the temperature dependence of resistance using a quantum
design PPMS (Physical Properties Measurement System) device. A
classical four-probe approach was applied. We measured the
resistances of samples in the magnetic fields of 0T, 0.5T, 1T, 2T,
4T, 6T and 8T, respectively, in the direction perpendicular to the
surface of the sample. Once the sample reached 50K, when it had
transferred from superconducting state to normal state completely,
the magnetic field was removed, and the sample was cooled in zero
field again to measure its resistance under another magnetic
field.
\par
ICP (Inductively Coupled Plasma Spectrometry) measurement was also
performed on several samples to obtain molar ratio of magnesium to
boron. To avoid the influence of extra magnesium attached to the
edge of the substrate, we carefully scratched the film from the
substrate to nitric acid in which the powder dissolved.

\section{Results and discussion}
The precursors were all around $2000\rm \AA$ thick, including the
Mg cap of around $180\rm \AA$. After annealing, the thin films
were around $4000 \rm \AA$ thick. The reason why the thin film
thickened after annealing can be explained as the result of
epitaxial growth, the oxidation of the magnesium cap, etc.
However, these explanations are not sufficient. We are still
seeking for the reason.
\par
\begin{figure}[p]
 \begin {center}
  \includegraphics[scale=1.3]{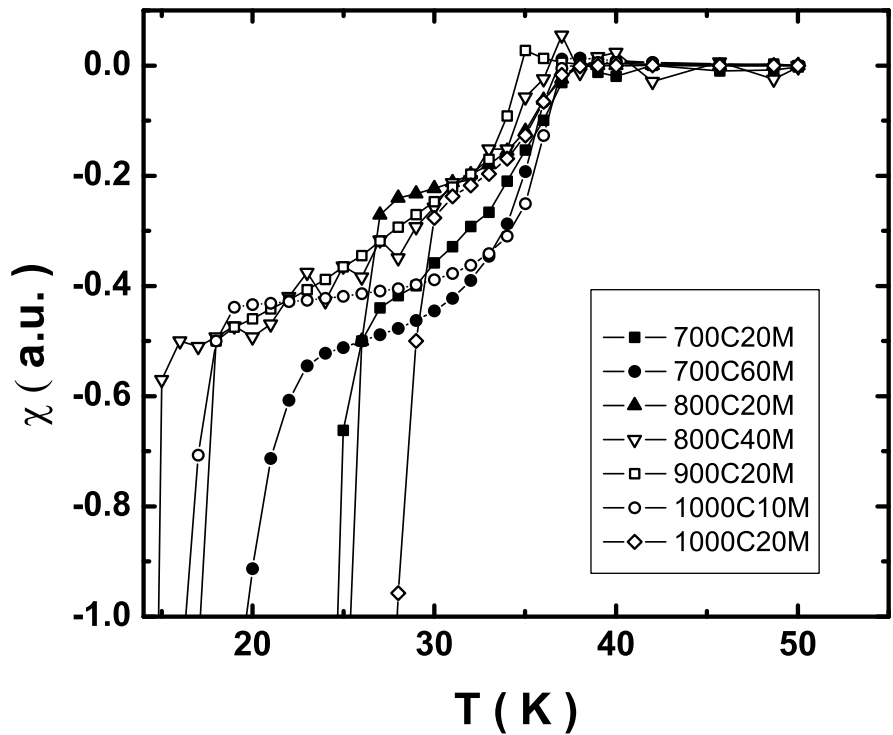}
 \end{center}
 \caption{Some typical two-transition normalized temperature dependence of
  magnetization. All these samples have a magnetization drop at
  around 38K, many of them have a secondary transition around 20K.}
 \label{fig1}
\end{figure}
\begin{figure}[p]
 \begin {center}
  \includegraphics[scale=1.3]{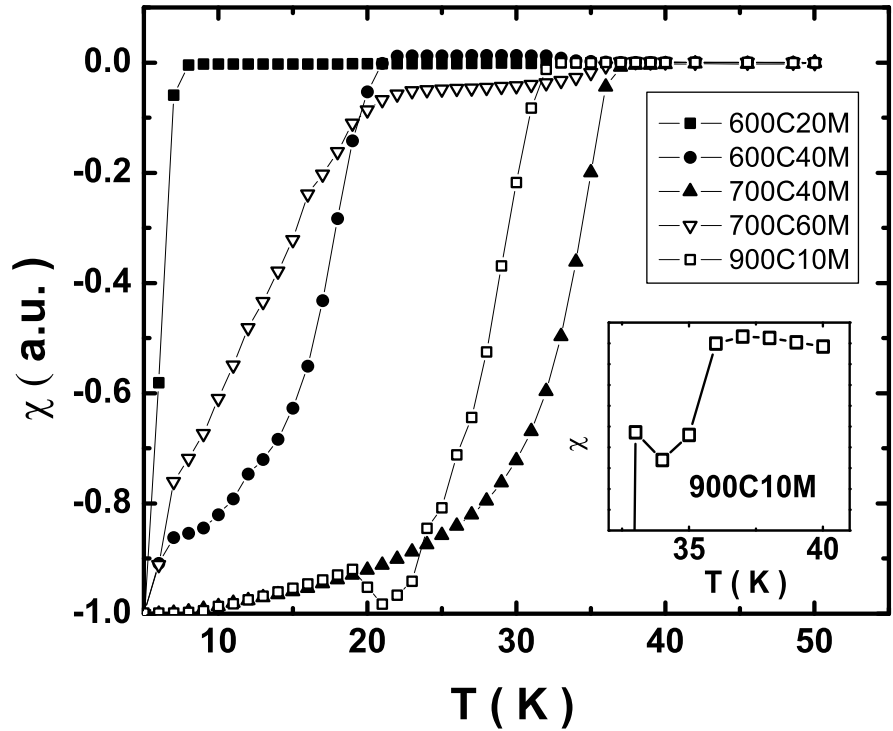}
 \end{center}
 \caption{Full chart for normalized temperature dependence of magnetization of several samples.
  Two transitions in sample 600C40M and 700C60M can be clearly observed. A tiny secondary
  transition of sample 900C10M is also shown in the inset.}
 \label{fig2}
\end{figure}

In magnetic measurement, we found the temperature dependence of
magnetization of most samples contains two transitions, as shown
in figure \ref{fig1} and figure \ref{fig2}. The distribution of
transition temperatures is not random but concentrates in several
certain temperatures: most of them have a magnetization drop at
around 38K, which is weaker compared to another magnetization drop
at around 20K. In figure \ref{fig2} we can see the transition
temperature of sample 600C40M is 22K. However, it has a secondary
transition at 8K, which is exactly the transition temperature of
sample 600C20M. As a typical two-transition
 sample, full chart of 700C60M is presented in figure \ref{fig2}.
 We can observe its two transitions clearly, and
 from the value of diamagnetization we can infer the superconductivity
 of 38K-T$_{c}$ compositions is
 much weaker than that of 22K-T$_{c}$ compositions. Both 700C40M
 and 900C10M have high transition temperatures. Actually 900C10M has
 a little magnetization drop at 36K
 before its primary transition at 33K, as the inset of figure \ref{fig2}
 shows. The uniform
existence of more than one magnetization transitions indicates
that these thin films contain several superconducting compositions
with different critical temperatures. Table \ref{tab1} summarized
the magnetic measurement results.
\par
\begin{figure}[p]
 \begin{center}
  \includegraphics[scale=1.3]{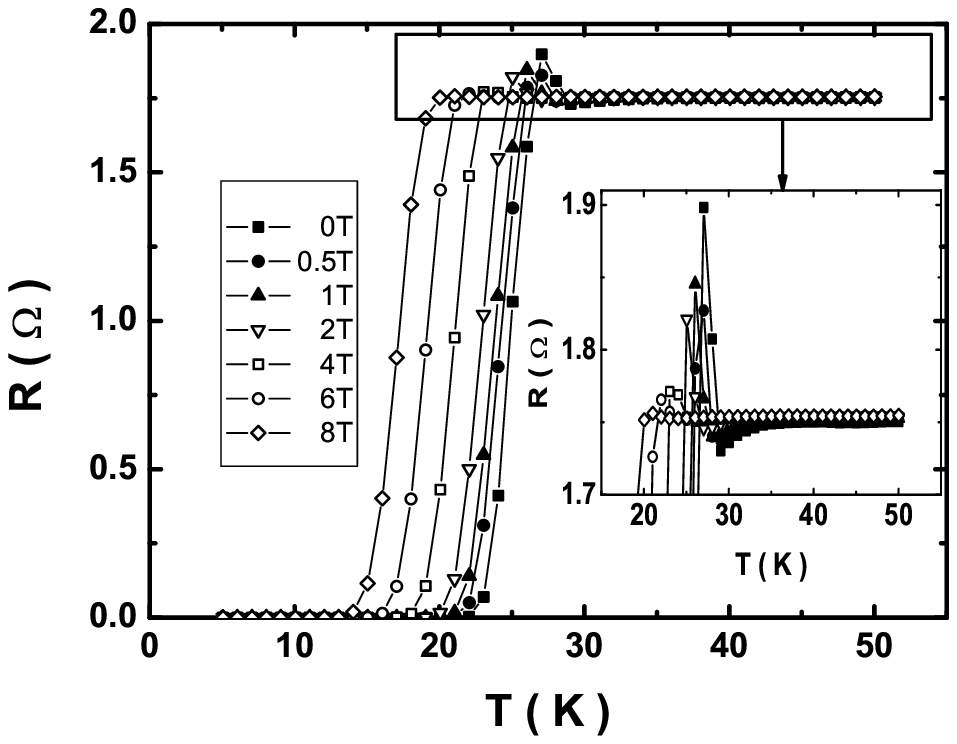}
 \end{center}
 \caption{Temperature dependence of the resistivity under different magnetic fields for sample 800C20M.}
 \label{fig6}
\end{figure}

More than one transitions in magnetic measurement is also reported
in the works of Shinde \emph{et al.}\cite{Shinde2001} and Ivanov
\emph{et al.}\cite{Ivanov2002}. Despite the common transition at
39K, Shinde \emph{et al.} reported transitions at 22K and 8K,
while Ivanov \emph{et al.} reported transition at around 20K. They
explained this phenomena as the deficiency of magnesium, which led
to the formation of MgB$_{4}$, MgB$_{6}$, etc. However, according
to our results of ICP analysis (Table \ref{ICP}), all samples were
excessively Mg-riched. Therefore the deficiency of magnesium may
not be able to explain the secondary transition. To the contrary,
sample 800C20M, which has the lowest Mg-B molar ratio, showed the
strongest diamagnetization around 38K among all samples analyzed
in ICP. Moreover, the temperature dependence of the resistance of
800C20M (figure \ref{fig6}) and 900C20M (figure \ref{fig8}) shows
the former one has much stronger superconductivity than the latter
one, which shows the highest Mg-B molar ratio in ICP analysis.
\par
\begin{figure}[p]
 \begin{center}
  \includegraphics[scale=1.3]{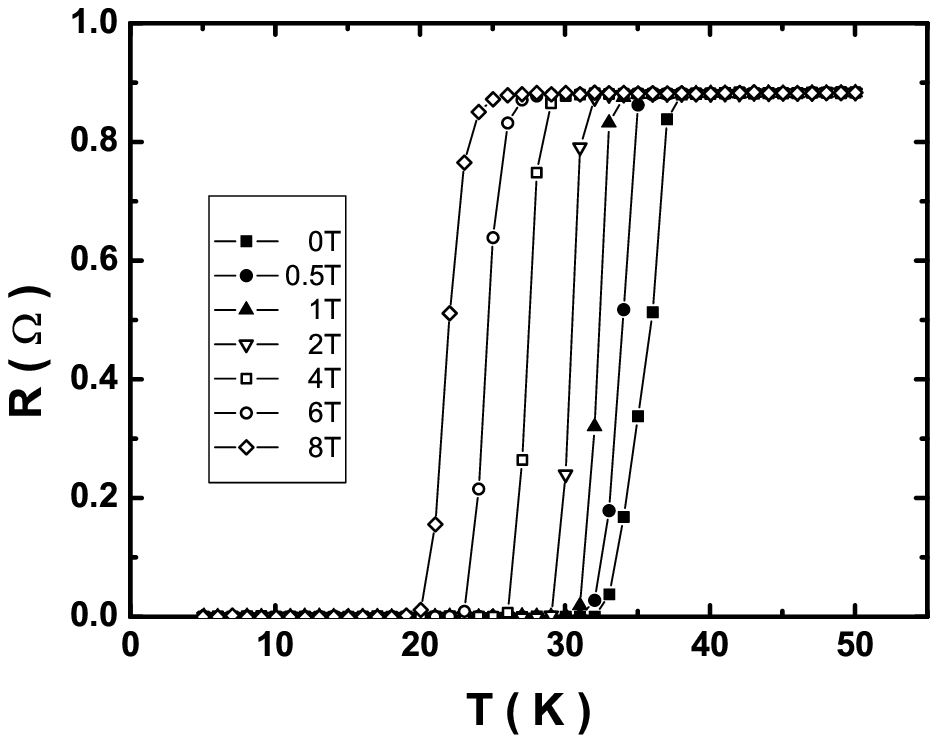}
 \end{center}
 \caption{Temperature dependence of the resistivity under different magnetic fields for sample
 900C10M.}
 \label{fig7}
\end{figure}
\begin{figure}[p]
 \begin{center}
  \includegraphics[scale=1.3]{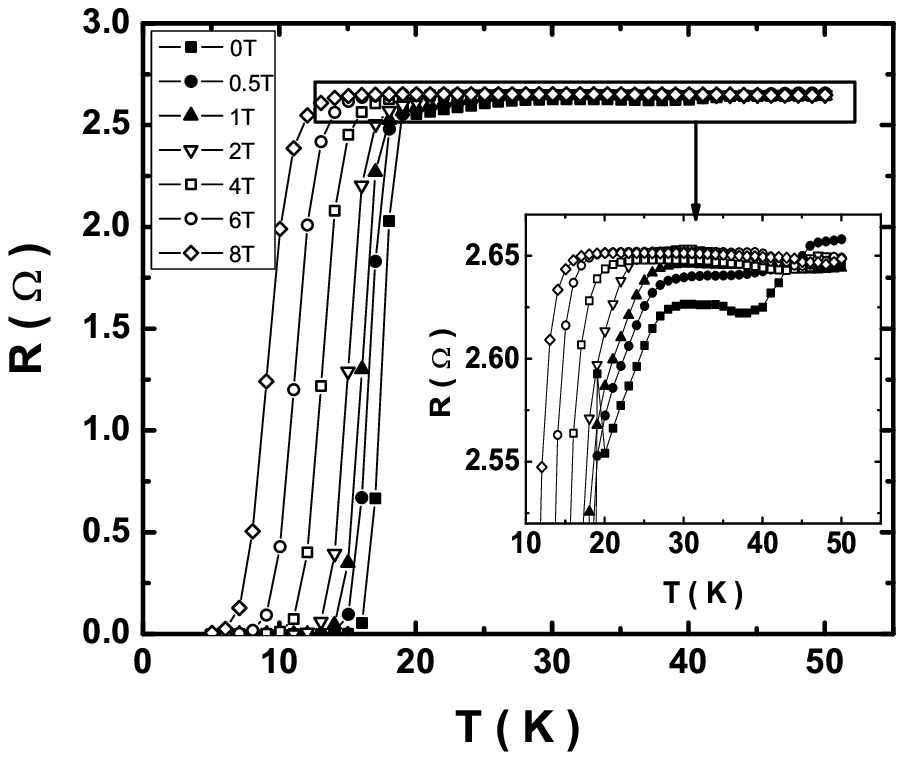}
 \end{center}
 \caption{Transition curves in different
 magnetic fields for sample
 900C20M.}
 \label{fig8}
\end{figure}
\begin{figure}[p]
 \begin{center}
  \includegraphics[scale=1.3]{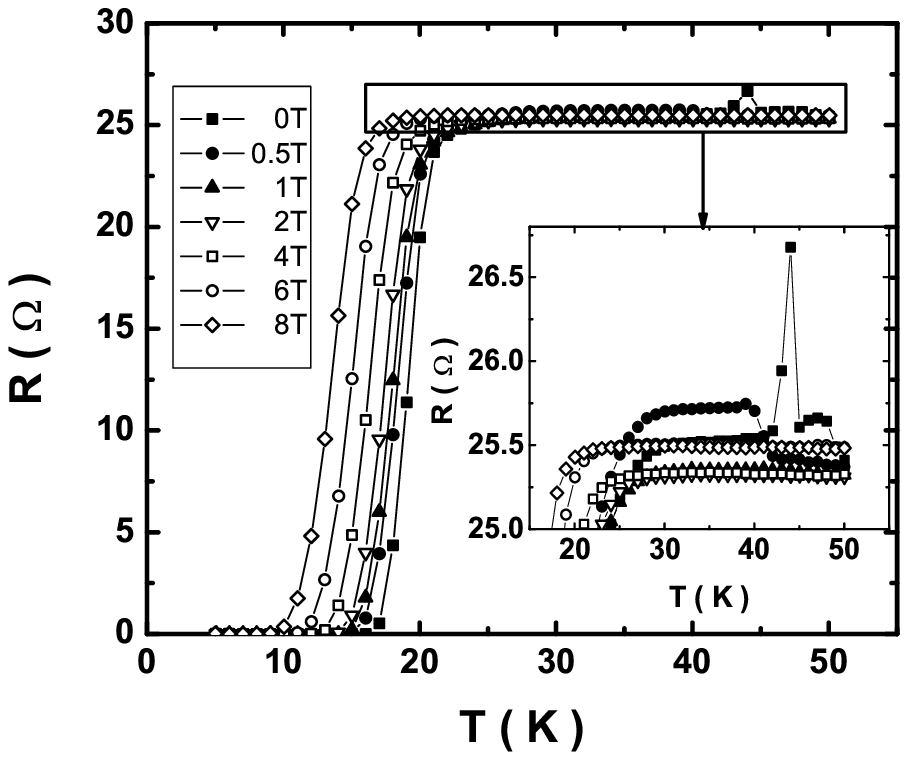}
 \end{center}
 \caption{Transition curves in different
 magnetic fields for sample
 1000C10M.}
 \label{fig9}
\end{figure}

The temperature dependence of the resistance of sample 800C20M,
900C10M, 900C20M and 1000C10M are shown in figure \ref{fig6},
\ref{fig7}, \ref{fig8} and \ref{fig9}, respectively. Strong
magnetic field has obvious effect to suppress the transition
temperature, which is also reported by others'
works\cite{Akimitsu2003}\cite{Drozd2004}\cite{Wang2004}. Sample
900C10M has the highest zero resistance temperature under any
field, while sample 900C20M has the lowest. Sample 1000C10M has
the similar zero resistance temperature as sample 900C20M does in
low field. However, strong magnetic field has greater effect on
sample 900C20M than that on sample 1000C10M.
\par
We can see that the zero resistance temperatures of the samples
under zero field are consistent with their lower transition
temperatures, not the upper ones, of their magnetic measurement
results. However, we can observe that the resistances start to
fall at around 38K under zero field, which matches their upper
transition temperatures, in the insets of figure \ref{fig6},
\ref{fig8} and \ref{fig9}. Considering the upper transitions at
around 38K in the results of magnetic measurement, the only
explanation to this fall is that some compositions of the thin
film turned to be superconducting under 38K, leading to local
short circuit and reduced the whole resistance. According to the
diamagnetization (see 700C60M in figure \ref{fig2}), only few
superconducting compositions have a T$_{c}$ of 38K. These
high-T$_{c}$ compositions are probable large superconducting
grains spread in some local areas of the film. The low zero
resistance temperature shows these high-T$_{c}$ compounds are not
topologically connected, so the resistance just falls, not
disappears at around 38K, although the sample has already shown a
diamagnetization. The sample turns to be completely
superconducting only after its lower transition, when most of its
compositions turn to be superconducting.
\par
The magnetic field we utilized in magnetic measurement is on the
different direction to that we utilized in resistance measurement,
but it does not weaken our comparison above. The magnetic field we
utilized in magnetic measurement was very low, and would have
little influence to the superconductivity of the film. Meanwhile
the temperature dependence of resistance we used for comparison
was obtained under zero field. Results of both measurements
reflected the inner properties of the sample, not the interaction
between the sample and the external magnetic field. Therefore the
difference direction of magnetic fields is not a problem here.
\par
M. Rajteri \emph{et al} explained such existence of two
transitions as "a diphasic percolation process"\cite{Rajteri2004}.
It was reasonable but a bit too simple. To give more detailed
explanations, we noticed the crystal lattice mismatch between the
sapphire substrate and the MgB$_{2}$ film on it. Such mismatch has
remarkable influence on the film, especially for the bottom layer
of the film\cite{Tian2002}. We have reason to believe the mismatch
lead to the lattice distortion of the bottom layer of the film.
Because density of states at $E_{F}$ of MgB$_{2}$ is very
sensitive to the lattice parameters\cite{Chen2003}, the change of
lattice parameters will affect the superconductivity of the bottom
layer greatly. We conclude that the transition temperature of the
bottom layer is different from that of the top layer, leading to
multi transitions. However, this is just a theoretical analysis
and we do not have direct supportive evidence, like SEM photo or
XRD. We are working to find out the exact substance of the
low-T$_{c}$ compositions.
\par
The discrepancy between temperature dependence of resistance and
temperature dependence of magnetization is not only reported in
our study. They are reported to be consistent in many
works\cite{bu2002}\cite{wnk2001}\cite{Li2001}. In the work of W.
N. Kang \emph{et al}\cite{wnk2003}, however, all their samples
showed a T$_{c}$ around 38K in resistive measurement, while most
samples showed a significant magnetization drop far below 38K in
their magnetic measurement. They did not give a detailed show of
temperature dependence of magnetization, but from their resistance
measurement we can infer that their samples all contained a few
compounds with T$_{c}$ around 38K. Different from our samples,
such compounds in theirs must have connected topologically so that
the resistance disappeared before a significant drop of
magnetization.
\par
We believe the discrepancy between temperature dependence of
resistance and temperature dependence of magnetization is the
result of competition of superconducting compounds with different
critical temperatures. If the high-T$_{c}$ compounds have
connected topologically, the resistance turns to be zero before a
significant drop of magnetization, otherwise the resistance will
still exist when a diamagnetization is shown.
\par
Multi transitions has never been reported in the papers of high
quality MgB$_{2}$ films, because the singularity of high-T$_{c}$
compound covered properties of other compounds. Our samples,
although were not of best quality, provide a perfect opportunity
to investigate some basic physical characteristics of this
material.
\par
Another noticeable phenomena in the temperature dependence of
resistance is the systematically sharp increase of resistance just
above the transition temperature. This cannot be explained as the
overshoot of current because more than one minute would be cost to
stabilize the system temperature, in which both the current and
the voltage had enough relaxation time, before the measurement of
resistance was taken. In sample 800C20M such resistance peak
appeared under any magnetic field (the inset of figure
\ref{fig6}), and the field had the effect to suppress the height
of the peak. In sample 1000C10M the peak appeared under weak
fields (0T and 0.5T)(the inset of figure \ref{fig9}), while in
sample 900C20M the peak only appeared under zero field (the inset
of figure \ref{fig8}). Sample 900C10M did not show this peak. The
peak may be the result of the competition and transformation
between superconductor and semiconductor compounds, but this
cannot explain why the magnetic field can suppress the height of
the peak. The research on this strange but systematic phenomena is
undergoing.
\par
\begin{figure}[p]
 \begin{center}
  \includegraphics[scale=1.3]{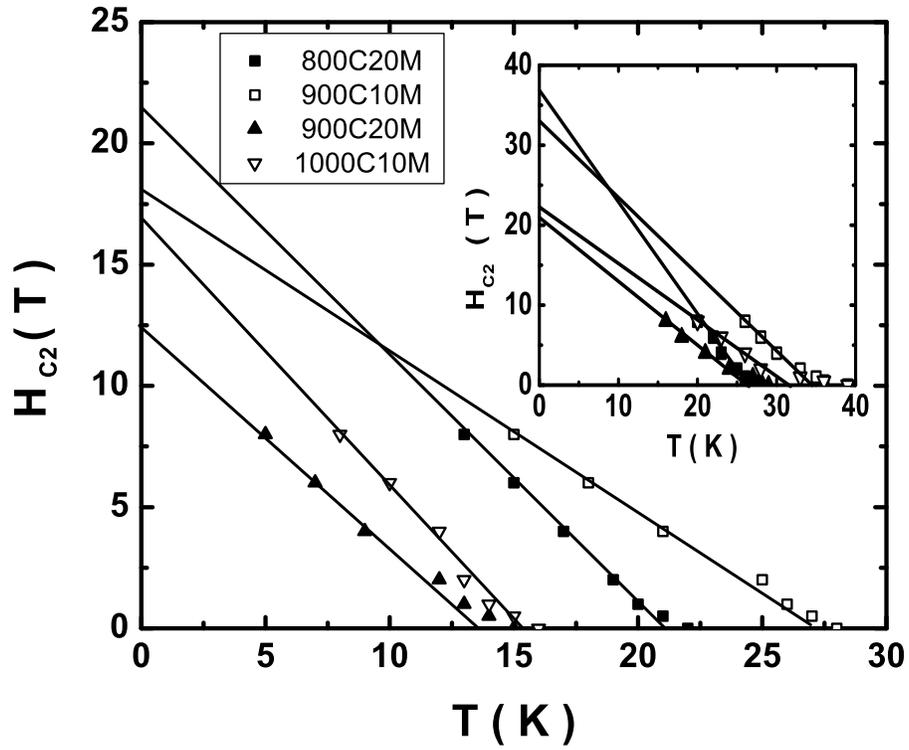}
 \end{center}
 \caption{Critical field based on zero resistance temperature.
 The magnetic field is perpendicular to the surface of the films. Critical
 field based on onset temperature is presented in the inset.}
 \label{fig10}
\end{figure}

On the basis of Ginzburg-Landau Theory, we can estimate the upper
critical fields of samples based on their transition curves in
different magnetic fields. Critical fields extracted are shown in
figure \ref{fig10} (based on the zero resistance temperature).
From the figure we can estimate sample 800C20M has the highest
upper critical field: 15.5T. Noticing the magnetic fields we
utilized are perpendicular to the films, this is a relatively high
value\cite{gurevich2004}. Sample 900C20M has the lowest critical
field: 8.8T. We noticed that sample 900C10M remains
superconducting up to around 20K in magnetic field of 8T,
revealing its capability to work under strong magnetic field at a
temperature that can be easily provided by modern cryocooler. We
also extracted the critical fields from the onset temperatures of
the samples, as shown in the inset of figure \ref{fig10}. Of
course it is much higher than the critical fields extracted from
zero resistance temperatures. Again sample 800C20M has the highest
value: 26.2T, while 900C20M still has the lowest value: 14.8T. We
have to point out this is just an approximate estimation. To
estimate more strictly, two-band G-L theory should be
utilized\cite{Askerzade2003}.

\section{Summary}
Precursors were prepared on Al$_{2}$O$_{3}$(0001) substrates by
codeposition of Mg and MgB$_{2}$ targets using pulsed laser
deposition technique. Superconducting MgB$_{2}$ thin films were
obtained via an \emph{ex situ} postannealing process, with various
annealing temperatures and durations. The superconducting
properties of the samples were intensively investigated. In
magnetic measurement, more than one transitions appeared in almost
all the samples, suggesting these films contained several
superconducting compounds. The temperature dependence of
resistance confirmed this phenomena. The ICP analysis proved this
is not result of magnesium deficiency. We believe the resistively
measured critical temperature of MgB$_{2}$ thin films is
determined by the competition of superconducting compounds with
different critical temperatures.

\section*{Acknowledgement}
This work was supported by Research Project (02002) of Educational
Ministry of China.

\begin{table}[p]
 \caption{The annealing conditions and magnetic measurement results of the
 samples.\newline
 (Annl.=Annealing, Temp.=Temperature, U.=Upper, L.=Lower, Trans.=Transition)}
 \label{tab1}
 \vspace{5mm}
 \begin{center}
  \begin{tabular}{l c c c c} \hline
    Sample ID&Annl. Temp.&Annl. Time&U. Trans. Temp.&L. Trans. Temp.\\
    \hline 600C20M&600$^{\rm o}$C&20min&8K&---\\
    600C40M&600$^{\rm o}$C&40min&22K&8K\\
    700C20M&700$^{\rm o}$C&20min&38K&22K\\
    700C40M&700$^{\rm o}$C&40min&38K&---\\
    700C60M&700$^{\rm o}$C&60min&38K&22K\\
    800C20M&800$^{\rm o}$C&20min&38K&26K\\
    800C40M&800$^{\rm o}$C&40min&38K&15K\\
    900C10M&900$^{\rm o}$C&10min&36K&33K\\
    900C20M&900$^{\rm o}$C&20min&35K&18K\\
    1000C10M&1000$^{\rm o}$C&10min&38K&18K\\
    1000C20M&1000$^{\rm o}$C&20min&38K&30K\\
    \hline
  \end{tabular}
 \end{center}
\end{table}

\clearpage
\begin{table}
 \caption{ICP analysis result}
 \label{ICP}
 \vspace{5mm}
 \begin{center}
  \begin{tabular}{l c c c}
  \hline
  Sample&\multicolumn{2}{c}{Concentration}&Molar Ratio\\
  ID&Magnesium&Boron&Mg:B\\
  \hline
  600C20M&2.48mg/L&0.332mg/L&3.42\\
  700C20M&3.64mg/L&0.418mg/L&3.99\\
  800C20M&3.66mg/L&0.550mg/L&3.05\\
  900C20M&5.24mg/L&0.371mg/L&6.47\\
  \hline
  \end{tabular}
 \end{center}
\end{table}

\clearpage

\end{document}